\documentclass[
 aip,
 amsmath,amssymb,
preprint
]{./revtex4-2}

\usepackage{natbib}
\usepackage{graphicx}
\usepackage{dcolumn}
\usepackage{bm}
\usepackage{multirow}
\usepackage{amsmath}
\usepackage{makecell}
\usepackage{mathrsfs}

\usepackage{tikz}

\begin{document}

\preprint{PoF/manuscript}

\title{A Swin-Transformer-based Model for Efficient Compression of Turbulent Flow Data}

\author{Meng Zhang} 
\email[]{These authors contributed equally to this work.}

\affiliation{School of Mechanical Engineering, Pusan National University, 2, Busandaehak-ro 63beon-gil, Geumjeong-gu, Busan, 46241, Republic of Korea}

\author{ Mustafa Z. Yousif}
\email[]{These authors contributed equally to this work.}
\affiliation{School of Mechanical Engineering, Pusan National University, 2, Busandaehak-ro 63beon-gil, Geumjeong-gu, Busan, 46241, Republic of Korea}
\affiliation{German Engineering Research and Development Center, LSTME-Busan Branch, 1276, Jisa-dong, Gangseo-gu, Busan, 46742, Republic of Korea}

\author{Linqi Yu}
\author{Hee-Chang Lim }
\email[]{Corresponding author, hclim@pusan.ac.kr}

\affiliation{School of Mechanical Engineering, Pusan National University, 2, Busandaehak-ro 63beon-gil, Geumjeong-gu, Busan, 46241, Republic of Korea}

\thanks{}

\begin{abstract}
This study proposes a novel deep-learning-based method for generating reduced representations of turbulent flows that ensures efficient storage and transfer while maintaining high accuracy during decompression. A Swin-Transformer network combined with a physical constraints-based loss function is utilized to compress the turbulent flows with high compression ratios and then restore the data with the underlying physical properties. The forced isotropic turbulent flow is used to demonstrate the ability of the Swin-Transformer-based (ST) model, where the instantaneous and statistical results show the excellent ability of the model to recover the flow data with remarkable accuracy. Furthermore, the capability of the ST model is compared with a typical Convolutional Neural Network-based auto-encoder (CNN-AE) by using the turbulent channel flow at two friction Reynolds numbers $Re_\tau$ = 180 and 550. The results generated by the ST model are significantly more consistent with the DNS data than those recovered by the CNN-AE, indicating the superior ability of the ST model to compress and restore the turbulent flow. This study also compares the compression performance of the ST model at different compression ratios ($\textit{CR}$) and finds that the model has low enough error even at very high $\textit{CR}$. Additionally, the effect of transfer learning (TL) is investigated, showing that TL reduces the training time by 64\% while maintaining high accuracy. The results illustrate for the first time that the Swin-Transformer-based model incorporating a physically constrained loss function can compress and restore turbulent flows with the correct physics.
\end{abstract}

\maketitle

\section{Introduction}\label{sec:introduction}
Turbulence, represented by the chaotic interactions among multiple spatial and temporal flow scales, has a significant impact on various fields such as aerospace \cite{Etkin1981}, environment \cite{Maxey1987}, wind energy \cite{Moriartyetal2002, Abdulqadiretal2017}, and combustion \cite{Kuo&Acharya2012}. With the development of measurement technologies and computing power, high-quality turbulence data can be obtained through experiments or simulations. In terms of experiments, hot-wire anemometry \cite{Bradbury1976, Bruun1996}, Particle Image Velocimetry (PIV) \cite{Westerweeletal2013}, and Particle-Tracking Velocimetry (PTV) \cite{Kasagi&Nishino1991} can measure the instantaneous velocity fields of turbulent flows with high accuracy and high spatial and temporal resolution. In terms of simulations, several computational fluid simulations are making it possible to process large amounts of data quickly and accurately, such as Reynolds-Averaged Navier-Stokes (RANS) models \cite{Alfonsi2009}, Large Eddy Simulation (LES) \cite{Moeng1984}, and Direct Numerical Simulation (DNS) \cite{Moin&Mahesh1998}. The advancement of experimental and simulation techniques and the increasing demand for high-quality turbulence data have led to large amounts of high-dimensional data, posing great challenges in storage and transmission. Therefore, efficient and accurate data compression techniques are necessary to reduce storage requirements, facilitate data transfer, and extract the main features of the flow field. Efficient storage and transmission methods are critical to turbulence research and help to understand the complex behavior of turbulence. 

Typically, data compression techniques extract the most critical features in the data while eliminating redundant or irrelevant information. Some techniques have been developed for the efficient storage and transfer of data. Singular value decomposition (SVD), a classic matrix decomposition technique, has been applied for data dimensionality reduction, feature extraction, and dynamic mode analysis\cite{Troppetal2017, Zimmermannetal2018}. Principal component analysis (PCA) (usually termed as proper orthogonal decomposition (POD) in the fluid dynamics community)\cite{Lumley1967, Kambhatla&Leen1997, Hubertetal2005, Serneels&Verdonck2008}, an unsupervised linear mapping compression method based on SVD technique, transforms the high dimensional data into the lower representation. Dynamic mode decomposition (DMD) is also based on SVD to compute the low-rank representation of the spatio-temporal flow data\cite{Kutz2017}. The above methods for data compression are all linear techniques, which makes them sensitive to outliers in the data. Another limitation of the above methods is they can not handle translation, rotation, and scaling of the data\cite{Kutz2017}. Furthermore, many nonlinear methods have been developed to capture complicated nonlinear structures in data. Kernel Principal Component Analysis (KPCA) was proposed by Schölkopf {\it et al.} \cite{Schölkopfetal2005}, which can efficiently compute principal components in high dimensional spaces by using integral operator kernel functions. Lee {\it et al.} \cite{Leeetal2004} compared two nonlinear projection algorithms, Isomap and Curvilinear Distance Analysis (CDA), and showed that Isomap is faster and theoretically more robust than CDA, while CDA is slower but more robust in practical applications. Hinton and Roweis \cite{Hinton&Roweis2002} introduced a probabilistic approach, called Stochastic neighbor embedding, for mapping high-dimensional representations or pairwise differences to a lower-dimensional space while preserving the neighborhood relations. A wavelet-based method incorporating a block-structured Cartesian mesh method was proposed by Sakai {\it et al.} \cite{Sakaietal2013} for the flow simulation data compression. Sifuzzaman {\it et al.} \cite{Sifuzzamanetal2009} compared the wavelet transform with the Fourier transform, revealing that the former approach took less response time. These methods provide more flexibility than linear compression methods but can result in high computation time and cost, especially for large datasets.

Thanks to big data, computing power, and algorithm development, machine learning has received extensive attention in recent decades and has been applied in various fields, such as computer vision \cite{Sebeetal2007, Zhangetal2018}, speech recognition \cite{Nodaetal2015}, natural language translation \cite{Collobertetal2011}, weather forecasting \cite{Hewageetal2021}, autonomous driving \cite{Grigorescuetal2020} and so on. In Fluid Dynamics, machine learning has been applied to solve several problems, such as flow denoising and reconstruction \cite{Fukamietal2019, Furutaetal2019, Liuetal2020, Kimetal2021, Yousifetal2021, Yousifetal2023a, Yousifetal2022a, Yuetal2022}, flow prediction \cite{Guastonietal2021, Lee&You2019}, active flow control \cite{Rabaultetal2019, Fanetal2020}, and turbulent inflow generation \cite{Yousifetal2022b, Yousifetal2023b}. The findings from the previous papers demonstrate the potential of deep learning to efficiently handle complex spatiotemporal data.  Furthermore, deep learning-based techniques have shown great promise over the past decades in compressing fluid flow data efficiently while preserving its main features. Liu {\it et al.} \cite{Liuetal2019} presented a data compression model using a generative adversarial network (GAN), where the discriminative network compresses data, and the generative network reconstructs data. They verified the performance of the GAN-based model on 3D flow past the cylinder, separation flow on the leeward of the double-delta wing, and shockwave vortex interaction. The results showed that the GAN-based model could save compression time and provide acceptable reconstruction quality. Glaws {\it et al.} \cite{Glawsetal2020} proposed a fully convolutional autoencoder deep-learning method to compress decaying homogeneous isotropic turbulence, Taylor-Green vortex, and turbulent channel flow. The study demonstrated the autoencoder model outperformed a variant of SVD with a similar compression ratio and had a good generalization. Furthermore, Olmo {\it et al.} \cite{Olmoetal2022} improved Glaws's work by leveraging the physical properties inherent in the CFD, which led to short training time and less training data under the same quality reconstructions. Yousif {\it et al.} \cite{Yousifetal2022b} applied a multiscale convolutional auto-encoder with a subpixel convolution layer (MSCSP-AE) to obtain the compact representation of the turbulent channel flow and used Long-Short-Term-Memory (LSTM) Network as a sequence learning model to predict the flow field over time scales. Their results showed that the MSCSP-AE could capture the crucial feature of the flow field and then feed the compressed data to LSTM to ensure the model predicts the key pattern of the flow. In the papers mentioned above, the compression models utilize stacked convolutional layers as the basis for their models, where finite-size filters capture the spatial correlation between neighborhood points, creating a more compact representation. 

The convolutional layer plays a vital role in deep learning due to its ability to capture adjacent spatial information and its non-linear approximation algorithm. However, convolutional layers rely on the kernel, or receptive field, which is limited to acquiring only local spatial correlations within the kernel field, making it challenging to recognize complex patterns \cite{Liangetal2021, Luetal2021}. The padding operation is one of the important parts of the convolutional layer, which is used to keep the feature map size the same as the original input. Still, it may cause artifacts at the edges of the input data, potentially affecting the model's performance in various applications, including turbulent boundary layer reconstruction cite{Yousifetal2023b}. Additionally, the convolutional layer was originally used to solve the pixel prediction and reconstruction in images, where pixels are distributed uniformly in a rectangular or square region. However, when processing the non-uniform flow data in fluid mechanics, the convolutional layer requires pre-processing it into a uniformly cartesian mesh, which is unrealistic \cite{Chenetal2022}. Moreover, the convolutional layer could lack flow details and consequently give wrong results for complex geometries \cite{Hu&Zhang2022}. 

 Recently, Transformer \cite{Vaswanietal2017} has achieved some success in sequence prediction and natural language processing (NLP) \cite{Radfordetal2018, Devlinetal2018, Wuetal2020, Zhouetal2021, Yousifetal2023b}, as its attention mechanism can discover the long-term dependencies in data, which has also sparked attention to its potential in computer vision applications. For example, Carion {\it et al.} \cite{Carionetal2020} introduced Detection Transformer (DETR) for objection detection. Dosovitskiy {\it et al.} \cite{Dosovitskiyetal2020} proposed the Vision Transformer (ViT) for image classification tasks and demonstrated that ViT outperforms CNNs. Han {\it et al.} \cite{Hanetal2021} proposed the Transformer in Transformer (TNT) for visual recognition tasks, demonstrating better preservation of local information than ViT. Liu {\it et al.} \cite{Liuetal2021} introduced the Swin Transformer with the shifted window scheme to address the window artifact problems encountered in the ViT model and found that the Swin Transformer achieves advanced performance on object detection and semantic segmentation.   Thanks to the impressive performance of the Swin Transformer, there are a large number of papers that utilized the Swin Transformer to tackle various vision problems. Liang {\it et al.} \cite{Liangetal2021} restored high-quality images from low-quality images using Swin Transformers as deep feature extraction blocks and convolutional layers as shallow feature extraction blocks. Liu {\it et al.} \cite{Liuetal2022} extended the Swin Transformer model from image recognition to video recognition and performed well on Kinetics-400, Kinetics-600, and Something-Something v2 benchmarks. Lu {\it et al.} \cite{Luetal2021} developed an Image Compression using the variational autoencoder (VAE) architecture and Swin Transformer. Their study indicated that the Swin Transformer model requires significantly fewer model parameters than other advanced methods such as CNN-based learnt image encoding. Inspired by the success of Swin Transformer-based models in the computer vision field, this study proposes an efficient Swin-Transformer (ST)-based model incorporating the physical properties of the flow field for turbulent data storage and transmission. The ST model does not use convolutional layers to avoid the limitations of convolutional layers, such as artifacts caused by padding operation, local spatial limitations caused by the finite-size kernel, and the inapplicability of non-uniform grid data.

The remainder of this paper is organized as follows. Section 2 introduces the methodology of compressing and decompressing flow data using the proposed ST model. The Direct numerical simulation (DNS) datasets used for training and testing the ST model are described in section 3. In section 4, the results from testing the ST model are discussed, and section 5 provides a summary of the conclusions drawn from this study.

\section{Methodology}\label{sec:Methodology}
\begin{figure}[htbp]
  \centerline{\includegraphics[scale=0.46]{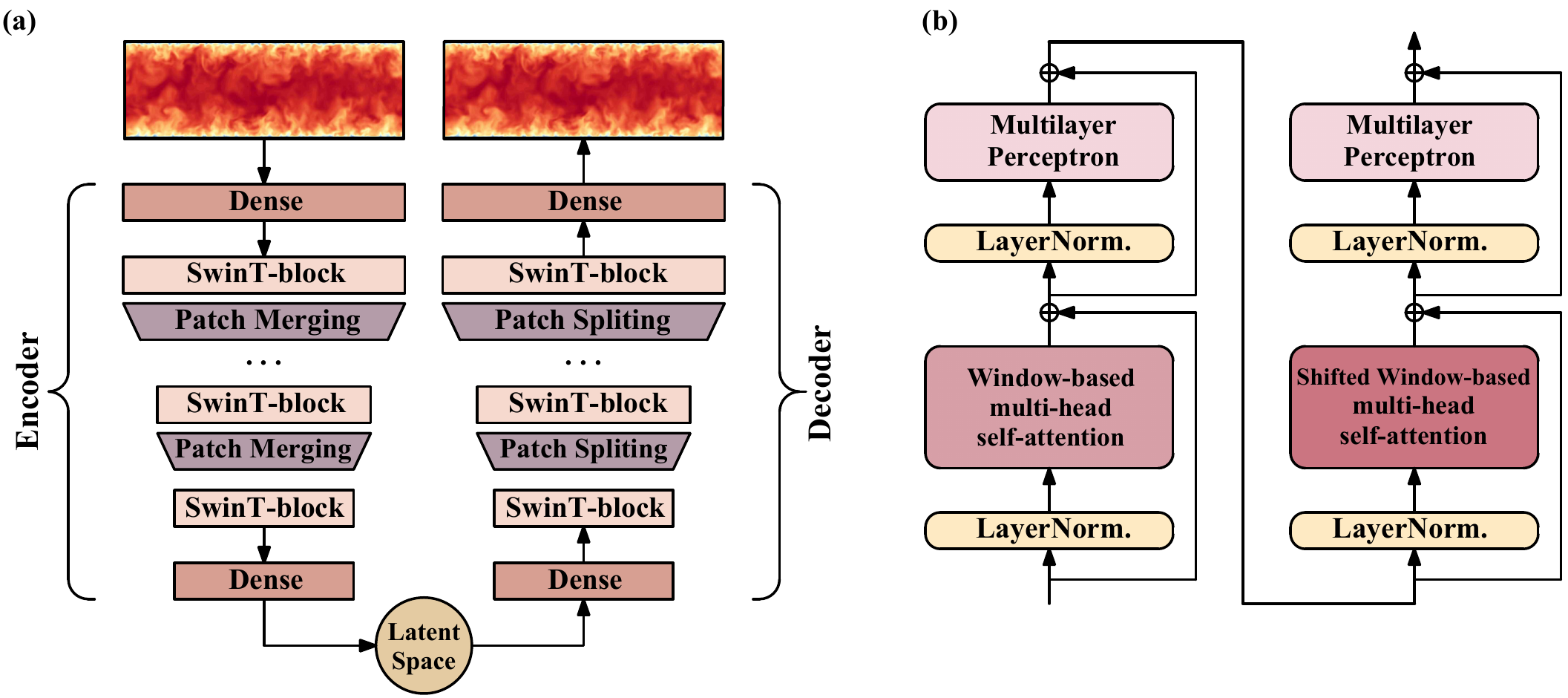}}
  \caption{The architecture of (a) the ST model and (b) the SwinT-block.}
\label{fig:F1}
\end{figure}

Transformer \cite{Vaswanietal2017} was originally proposed for NLP problems, but the ViT \cite{Dosovitskiyetal2020} adapted it for computer vision by splitting input images into patches, similar to NLP tokens. Therefore, the correlation between patches can be captured through the self-attention operation in Transformer, addressing the limitation of CNN kernels in capturing only local information. Swin Transformer \cite{Liuetal2021} improves upon the ViT model and incorporates shifted windows to avoid window artifact issues. The proposed ST model is based on Swin Transformer, which divides the input flow field data into multiple patches, groups them into several windows, and employs shifted windows to overcome the lack of window boundary information. The architecture of the ST model is shown in Figure~\ref{fig:F1} (a). The model consists of an encoder and a decoder. The encoder plays a critical role in reducing the input data size for efficient storage and transmission while maintaining the important features. The decoder is responsible for restoring the original data from the reduced representations with high accuracy. Figure~\ref{fig:F1} (a) shows that the encoder starts and ends with a dense layer, with a series of Swin Transformer blocks (SwinT-blocks) and patch-merging sandwiched in between. The decoder structure is symmetrical with the encoder one, but the patch-splitting replaces the patch-merging. Here, the dense layers at the beginning project the data to an arbitrary dimension $\textit{C}$, while the dense layers at the end project the data dimension back to the original dimension. The SwinT-block captures the main features of the data, which will be described in detail later. The patch-merging operation performs a similar function to the downsampling layer in CNN, which reduces the number of patches as the network is stacked. While the patch-splitting operation can be considered an upsampling layer, increasing the number of patches. It is worth noting that the entire architecture has no convolutional layers.

\begin{figure}[htbp]
  \centerline{\includegraphics[scale=0.45]{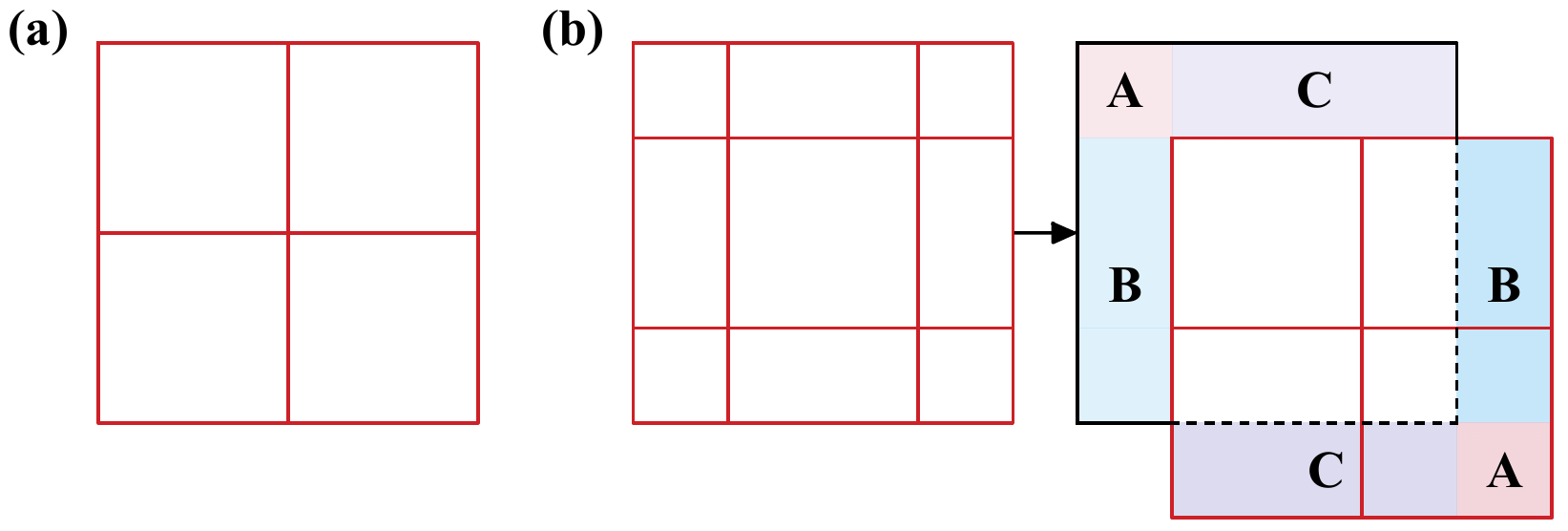}}
  \caption{The window partitioning method for (a) W-MSA and (b) SW-MSA. Here, each red block means one window to calculate the local self-attention.}
\label{fig:F2}
\end{figure}

As shown in Figure~\ref{fig:F1} (b), the SwinT-block mainly consists of a Window-based multi-head self-attention (W-MSA) and a Shifted Window-based multi-head self-attention (SW-MSA), both of them followed by a Multilayer Perceptron (MLP). Each W-MSA, SW-MSA, and MLP in the block is placed with a LayerNorm layer at the beginning, followed by residual connections that connect the output with its input. The ViT uses global self-attention to calculate relationships between all tokens, which increases the computational cost when the number of tokens is very large. However, unlike global self-attention in ViT, as Figure~\ref{fig:F2} (a) shows, the ST model uses local self-attention to compute self-attention within each non-overlapping local window, where each window contains $\textit{M×M}$ patches (with $\textit{M}$ set to 8 in this study). The computational complexity $\Omega$ of the global multi-head self-attention (MSA) and window-based MSA for input data of $\textit{h×w}$ size can be expressed as follows:
\begin{equation} \label{eqn:eq1}
\Omega (MSA) = 4hwC^{2} + 2(hw)^{2}C,
\end{equation}
\begin{equation} \label{eqn:eq2}
\Omega (W-MSA) = 4hwC^{2} + 2M^{2}hwC,
\end{equation}
\noindent here, the only difference is the last term, where the global MSA is quadratic to the input size ($\textit{hw}$), whereas the W-MSA is linear to $\textit{hw}$ when the value of $\textit{M}$ is fixed. Therefore, W-MSA is more cost-effective, especially for larger input sizes.

Furthermore, the lack of cross-window information, that is the connection on the boundaries of each window can be solved by using a shifted window multi-head self-attention (SW-MSA). The shifted window partitioning method cyclically shifts the divided window towards the upper-left direction to form a new window division with the same number of windows, as shown in Figure~\ref{fig:F2} (b). Then masking mechanism restricts self-attention from calculating non-adjacent window features.

Self-attention in W-MSA and SW-MSA is a function that maps a query and a set of key-value pairs to an output, and its formula is as follows:
\begin{equation} \label{eqn:eq3}
\textbf{\textit{Q}} = \textbf{\textit{XW}}_\textbf{\textit{Q}},
\end{equation}
\begin{equation} \label{eqn:eq4}
\textbf{\textit{K}} = \textbf{\textit{XW}}_\textbf{\textit{K}},
\end{equation}
\begin{equation} \label{eqn:eq5}
\textbf{\textit{V}} = \textbf{\textit{XW}}_\textbf{\textit{V}},
\end{equation}
\begin{equation} \label{eqn:eq6}
Attention(\textbf{\textit{Q,K,V}}) = SoftMax(\frac{\textbf{\textit{QK}}^{\top}}{\sqrt{d}} + \textbf{\textit{B}})\textbf{\textit{V}},
\end{equation}
\noindent where $\textbf{\textit{W}}_\textbf{\textit{Q}}$, $\textbf{\textit{W}}_\textbf{\textit{K}}$, $\textbf{\textit{W}}_\textbf{\textit{V}}$ are the weight matrices shared among all windows; $X \in \mathbb{R}^{M^{2}\times C}$ is one of the local window features, $\textbf{\textit{Q}}$, $\textbf{\textit{K}}$, $\textbf{\textit{V}} \in \mathbb{R}^{M^{2}\times d}$ are query, key and value matrices, respectively; $d$ is the dimension of query; $\textbf{\textit{B}} \in \mathbb{R}^{M^{2}\times M^{2}}$ is the learnable relative positional encoding. The attention function mentioned above is typically calculated multiple times, with the number of calculations equal to the number of attention heads used (referred to as $h$). The output of each attention calculation is then spliced together to form the final multi-head attention output.

The proposed ST model in this study incorporates physical principles to guide its learning process, facilitating the capture of the underlying physical behavior of turbulent flow and achieving better fitting to the training data. The first physical loss employed in the proposed ST model is the gradient error loss $L_{gradient}$, which is computed from the gradient of the flow. This loss term can assist the model in accurately reconstructing the turbulent flow with non-uniform grid distribution, particularly in the wall-normal direction of turbulent channel flow in this study. Reynolds stress error $L_{Reynolds~stress}$ and the spectrum error $L_{spectrum}$ quantify the variance in the Reynolds stress tensor of velocity fields and the difference in the spectral content of the flow parameters, respectively. By incorporating these loss terms, the model's ability to reconstruct the Reynolds stress components and the energy spectra of the flow is enhanced. In addition, the reconstructed velocity field error $L_{velocity}$ also be considered as the basic loss in this model. The loss functions for the proposed ST model are defined as follows:
\begin{equation} \label{eqn:eq7}
L_{gradient} = \frac{1}{S} \sum_{s=1}^{S} \lVert \nabla \hat{\textbf{\textit{x}}}_{s} - \nabla \textbf{\textit{x}}_{s} \lVert_2^2,
\end{equation}
\begin{equation} \label{eqn:eq8}
L_{Reynolds~stress} = \frac{1}{S} \sum_{s=1}^{S} \lVert \hat{\textbf{\textit{T}}}_{s} - \textbf{\textit{T}}_{s} \lVert_2^2,
\end{equation}
\begin{equation} \label{eqn:eq9}
L_{spectrum} = \frac{1}{S} \sum_{s=1}^{S} \lVert \hat{E}(k)_{s} - E(k)_{s} \lVert_1,
\end{equation}
\begin{equation} \label{eqn:eq10}
L_{velocity} = \frac{1}{S} \sum_{s=1}^{S} \lVert  \hat{\textbf{\textit{x}}}_{s} -  \textbf{\textit{x}}_{s} \lVert_2^2,
\end{equation}
\begin{equation} \label{eqn:eq11}
L_{total} = \lambda_{1}L_{gradient} + \lambda_{2}L_{Reynolds~stress} + \lambda_{3}L_{spectrum} + \lambda_{4}L_{velocity},
\end{equation}
\noindent where the quantities with " $\hat{~}~$" are the outputs of the ST model; $\lVert \cdot \lVert_1$ and $\lVert \cdot \lVert_2$ are the $L_1$ and $L_2$ norms; $\textbf{\textit{T}}$ expresses the Reynolds stress tensor; $E(k)$ is the energy spectrum, $k$ is the wavenumber; $S$ is batch size. The balance coefficients of the loss terms, denoted as $\lambda_{1}$, $\lambda_{2}$, $\lambda_{3}$ and $\lambda_{4}$, have been empirically determined as 0.01, 80, $10^{-5}$, and 300 for isotropic turbulent flow, respectively. For turbulent channel flow, they are set as 5, 100, $10^{-5}$, and 200, respectively.

\section{Data description and pre-processing}\label{sec:Data description and pre-processing}
In this study, we investigate two different types of flows: the forced isotropic turbulence flow obtained from the Johns Hopkins turbulence databases (JHTDB), which serves as a demonstration case, and the turbulent channel flow at $Re_\tau$ = 180 and 550 generated by performing DNS, which is used as systematic model capability test case. In both cases, the ST model is trained using an adaptive moment estimation (Adam) optimization algorithm \cite{Kingma&Ba2014} with a batch size $S$ = 8 and an initial learning rate $\eta$ = 0.0001. To implement the model, the open-source library TensorFlow 2.2.3 is utilized. Additionally, an early stopping regulation technique is employed to terminate the training.

\subsection{Forced isotropic turbulence flow data}
For the demonstration case, the forced isotropic turbulence dataset obtained from the JHTDB at a Taylor-scale Reynolds number $Re_\lambda=\lambda u_{rms}/\nu$ = 418 is considered to train and test the proposed ST model, where $\lambda=(15\nu u_{rms}^{2}/\varepsilon)^{1/2}$ is Taylor microscale, $u_{rms}=(\langle u_{i} u_{i} \rangle/3)^{1/2}$ represents root-mean-squared velocity, $\nu$ is the kinematic viscosity and $\varepsilon$ means dissipation rate. This dataset was generated from DNS using a pseudo-spectral parallel code. The governing equations used for simulation were the incompressible Navier-Stokes equations. The velocity vector $\textbf{\textit{u}}$ = ($u$, $v$, $w$), where $u$, $v$, $w$ are streamwise, wall-normal, and spanwise components, respectively, with the corresponding directions $x$, $y$, $z$. The grid points are uniformly distributed in all directions. The detailed parameters for the forced isotropic turbulence are shown in Table~\ref{tab:Table1}. Further information regarding the simulation and the database utilized in this study can be found in Perlman {\it et al.} \cite{Perlmanetal2007}.
\begin{table}[htbp]
  \begin{center}
\scalebox{1.0}{
\begin{tabular}{ccccccccc} \hline\hline
&$Re_{\lambda}$~ ~&~~$L_x\times L_y\times L_Z$~ ~&~~$N_x\times N_y \times N_z$~ ~&~~$\nu$~ ~&~~$\Delta t$&  \\ \hline
&$418$~ ~&~~$2 \pi\times 2 \pi\times 2 \pi$~~&~~$1024\times 1024\times 1024$~~&~~$0.000185$~~&~~$0.0002$&   \\
 \hline\hline
\end{tabular}}
  \caption{The detailed parameters for the forced isotropic turbulence. Here, $L$ is the domain dimension and $N$ is the number of grid points. $\nu$ and $\Delta t$ represent kinematic viscosity and simulation time-step, respectively.}
  \label{tab:Table1}
  \end{center}
\end{table}

The velocity dataset is applied as input to the ST model, which contains 200 snapshots of the $x-y$ plane (where $z$ = 0). The dataset spans approximately two large-eddy turnover times. The training dataset consists of 100 snapshots, and the test dataset is another 100 snapshots that are completely separate from the training dataset. The time interval between each snapshot in the training and testing dataset is 0.02. In order to reduce computational costs, the entire domain is divided into 64 parts, resulting in a change in data size from the original $N_x\times N_y=1024\times 1024$ in the $x-y$ plane to 128$\times$128. Consequently, the training dataset comprises 6400 sub-snapshots, which are randomly shuffled before being fed into the model.  

\subsection{Turbulent channel flow}
The turbulent channel flow data at $Re_\tau$ = 180 and 550 are utilized as datasets for the proposed model. The flow data are produced through DNS using the incompressible momentum and continuity equations, which are expressed as:
\begin{equation} \label{eqn:eq12}
\frac{\partial {\textbf{\textit{u}}}}{\partial t} + \nabla \cdot ({\textbf{\textit{u}}}{\textbf{\textit{u}}}) = -\frac{1}{\rho}\nabla p + {\nabla} \cdot ( \nu {\nabla} {\textbf{\textit{u}}}),
\end{equation}
\begin{equation} \label{eqn:eq13}
\nabla \cdot {\textbf{\textit{u}}} = 0.
\end{equation}

In the equations above, $\textbf{\textit{u}}$ = ($u$, $v$, $w$) denotes the velocity vector, where $u$, $v$ and $w$ represent the streamwise, wall-normal and spanwise components in $x$, $y$, $z$ directions. $t$, $\rho$, $p$, and $\nu$ are time, density, pressure, and kinematic viscosity, respectively. The open-source computational fluid dynamics (CFD) finite-volume code OpenFOAM-5.0x is used to perform the simulations.

The simulation parameters of each friction Reynolds number are shown in Table~\ref{tab:Table2}. The streamwise and spanwise directions are subject to periodic boundary conditions. Meanwhile, the channel top and bottom are subject to no-slip conditions. The grid points are uniformly distributed in the $x$ and $z$ directions, while a non-uniform distribution is used in the $y$ direction. DNS data obtained from Moser {\it et al.} \cite{Moseretal1999} have been used to validate the turbulence generated by the simulation, and it was verified that the simulated data had similar statistical characteristics. The simulation uses the pressure implicit split operator algorithm to solve the coupled pressure-momentum system. A second-order accurate linear upwind scheme is utilized to discretize the convective fluxes. Similarly, all other discretization schemes used in the simulation also have second-order accuracy.
\begin{table}
  \begin{center}
\scalebox{1.0}{
\begin{tabular}{cccccccccc} \hline\hline
&$Re_\tau$~ ~&~~$L_x\times L_y \times L_z$~ ~&~~$N_x\times N_y \times N_z$~ ~&~~$\Delta x^+$~ ~&~~$\Delta z^+$~ ~&~~$\Delta y_w^+$~ ~&~~$\Delta y_c^+$~ ~&~~$\Delta t^+$&  \\ \hline
&$180$~ ~&~~$4\pi\delta\times2\delta\times2\pi\delta$~~&~~$256 \times 128 \times 256$~~&~~$8.831$~~&~~$4.415$~~&~~$0.63$~~&~~$4.68$~~&~~$0.113$&   \\
&$550$~ ~&~~$4\pi\delta\times2\delta\times2\pi\delta$~~&~~$512 \times 336 \times 512$~~&~~$13.492$~~&~~$6.746$~~&~~$0.401$~~&~~$5.995$~~&~~$0.030$& \\ 
\hline\hline
\end{tabular}}
  \caption{Simulation parameters of turbulent channel flow at $Re_\tau$ = 180 and 550. Here, $L$ is the domain dimension and $N$ is the number of grid points. The superscript "$+$" denotes that the quantity is made dimensionless by using $u_\tau$ and $\nu$. $\Delta y_w^+$ refers to the distance near the wall and $\Delta y_c^+$ refers to the spacing in the center of the channel.}
  \label{tab:Table2}
  \end{center}
\end{table}

The training dataset contains 16,000 snapshots of a single ($y-z$) plane extracted from turbulent channel flow simulation, split evenly between turbulence data at $Re_\tau$ = 180 and $Re_\tau$ = 550, with 8,000 snapshots in each subset. Additionally, the test dataset for each case consists of another 1,000 snapshots. To apply transfer learning to the data at $Re_\tau$ = 550 by initializing the model weights with the weights of the flow at $Re_\tau$ = 180, we interpolate the data $Re_\tau$ = 550 to match the grid size of the data at $Re_\tau$ = 180. The interval between the collected snapshots of the flow fields equals ten simulation time steps for the flow at each $Re_\tau$.

\section{Results and discussion}
\subsection{Forced isotropic turbulence flow}
\begin{figure}[htbp]
  \centerline{\includegraphics[scale=0.6]{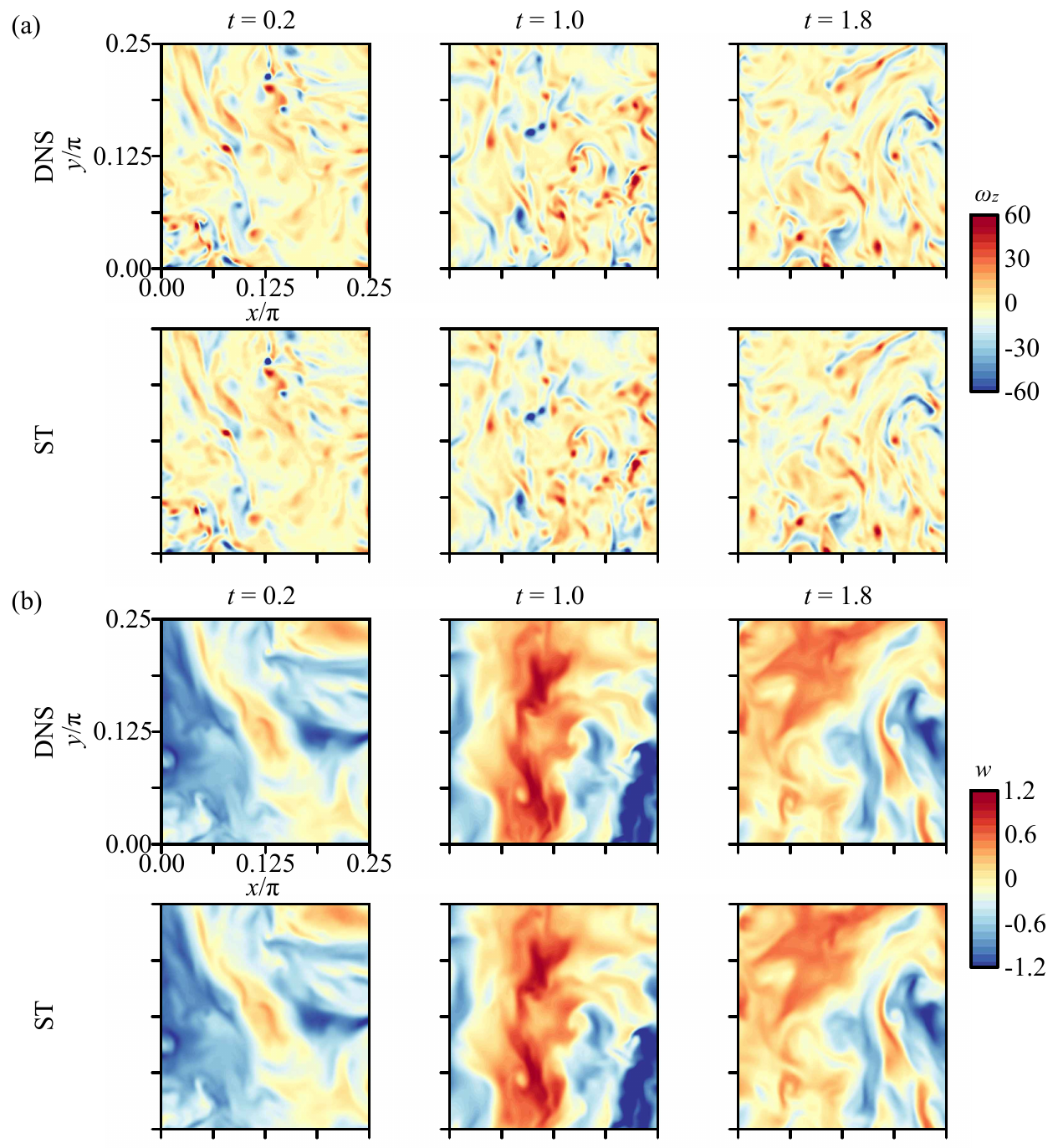}}
  \caption{Instantaneous spanwise (a) vorticity field and (b) velocity field for the case of forced isotropic turbulence.}
\label{fig:F3}
\end{figure}

In this section, the forced isotropic turbulence data are used to examine the ability of the ST model to compress and reconstruct data. The compression ratio ($CR$) is used to quantify the degree of compression achieved by the given model, where $CR$ = (original data size / compressed data size) (with $CR$ is 16 in this section). Additionally, test data that are not contained in the training are used to obtain subsequent results. The decompressed instantaneous spanwise vorticity field ($\omega_z$) and velocity field ($w$) for three different time steps are shown in Figure~\ref{fig:F3}. As can be observed, the ST model achieves a satisfactory qualitative reproduction of the true fields.

\begin{figure}[htbp]
  \centerline{\includegraphics[scale=0.32]{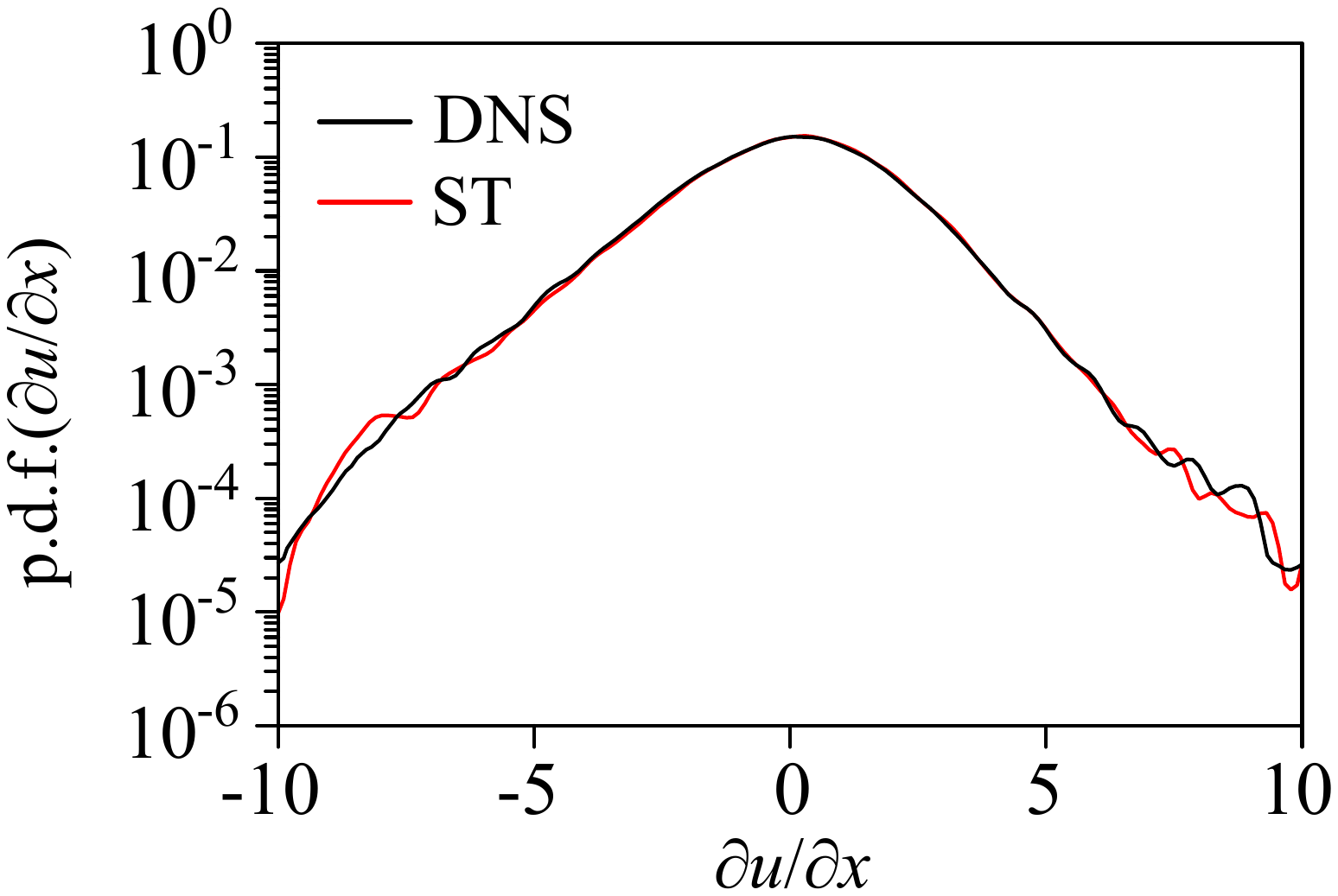}}
  \caption{Probability density function plot of the velocity gradient field for the case of isotropic turbulent flow.}
\label{fig:F4}
\end{figure}

\begin{figure}[htbp]
  \centerline{\includegraphics[scale=0.25]{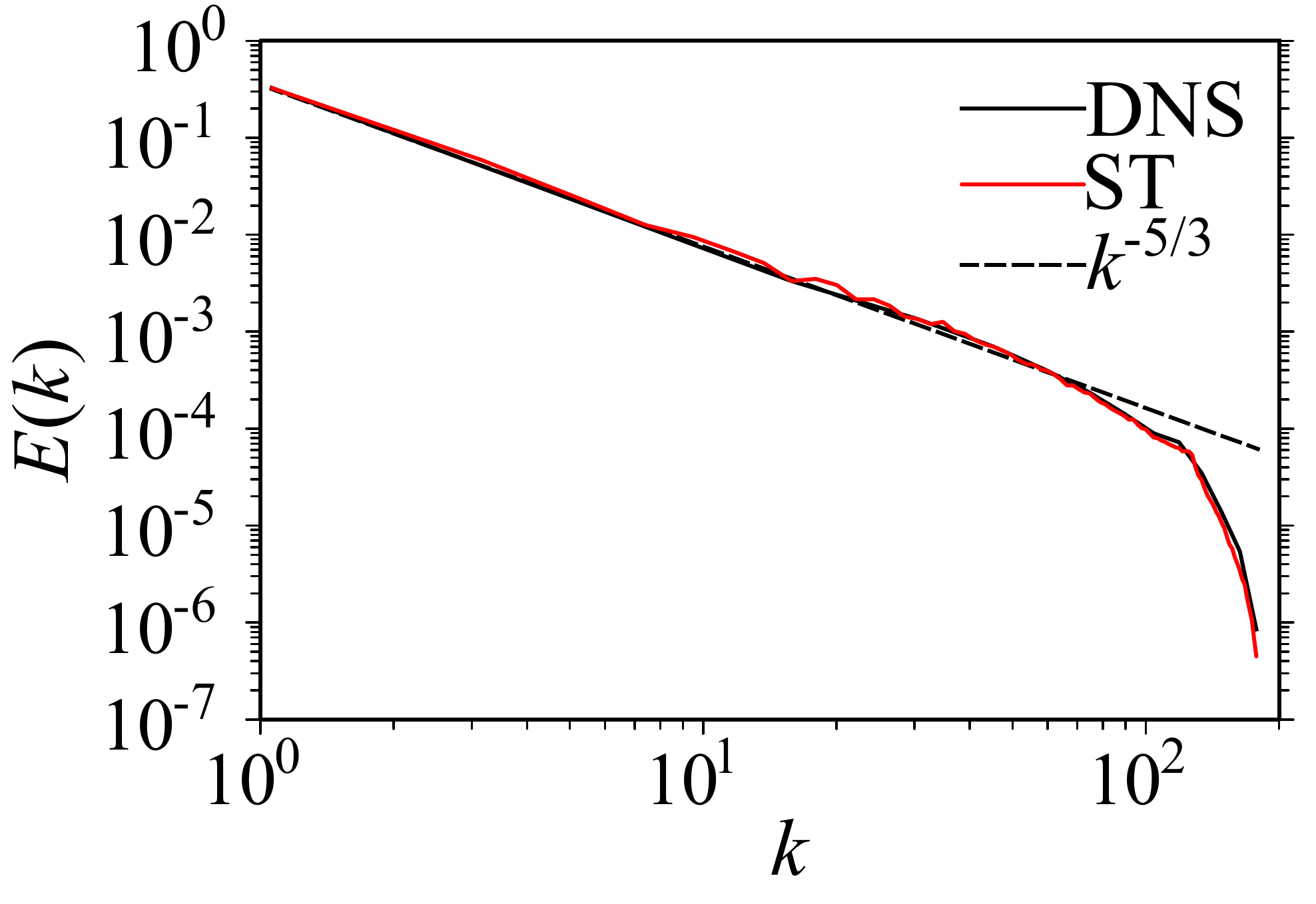}}
  \caption{Kinetic energy spectrum for the case of isotropic turbulent flow.}
\label{fig:F5}
\end{figure}

In addition to qualitative assessments, a detailed analysis of flow statistics is conducted to evaluate the performance of the ST model. Figure~\ref{fig:F4} displays the probability density function (p.d.f.) plot of the decompressed velocity gradient field ($\partial u/\partial x$), which demonstrates the ability of the ST model to accurately reconstruct the flow field. It is worth noting that slight deviations are shown at the tails of the p.d.f. because the decompressed flow fields are less intermittent than DNS data.  Furthermore, the Kinetic energy spectrum ($E(k)$) is used to check the performance of the ST model in terms of the inertial scale, where $k$ is the wave number. As shown in Figure~\ref{fig:F5}, the spectrum of the decompressed data agreed well with the DNS result, indicating that the ST model can reproduce the flow with an accurate spectrum content along the whole inertial scales.

There is clear evidence from the above demonstration results that the ST model is capable of compressing and decompressing the uniformly distributed turbulent flow effectively and maintaining the same instantaneous and statistical results as the ground truth data. In the next section, the ability of the ST model to reconstruct the non-uniformly distributed turbulent flow is verified.

\subsection{Turbulent channel flow}
In this section, the compression and decompression capabilities of the ST model are verified using turbulent channel flow at $Re_\tau$ = 180 and $Re_\tau$ = 550. To establish a baseline for comparison, the channel flow snapshots were compressed and reconstructed using a CNN-based autoencoder (CNN-AE) with an architecture similar to the ST model. Here, convolutional layers, downsampling, and upsampling are used instead of SwinT-blocks, patch-merging, and patch-splitting. Both the ST model and the CNN-AE have the same $CR$ of 64 and the same hyperparameters. In addition, this section evaluates the performance of the ST model at different $CR$, verifying the robustness of the model.

\begin{figure}[htbp]
  \centerline{\includegraphics[scale=0.4]{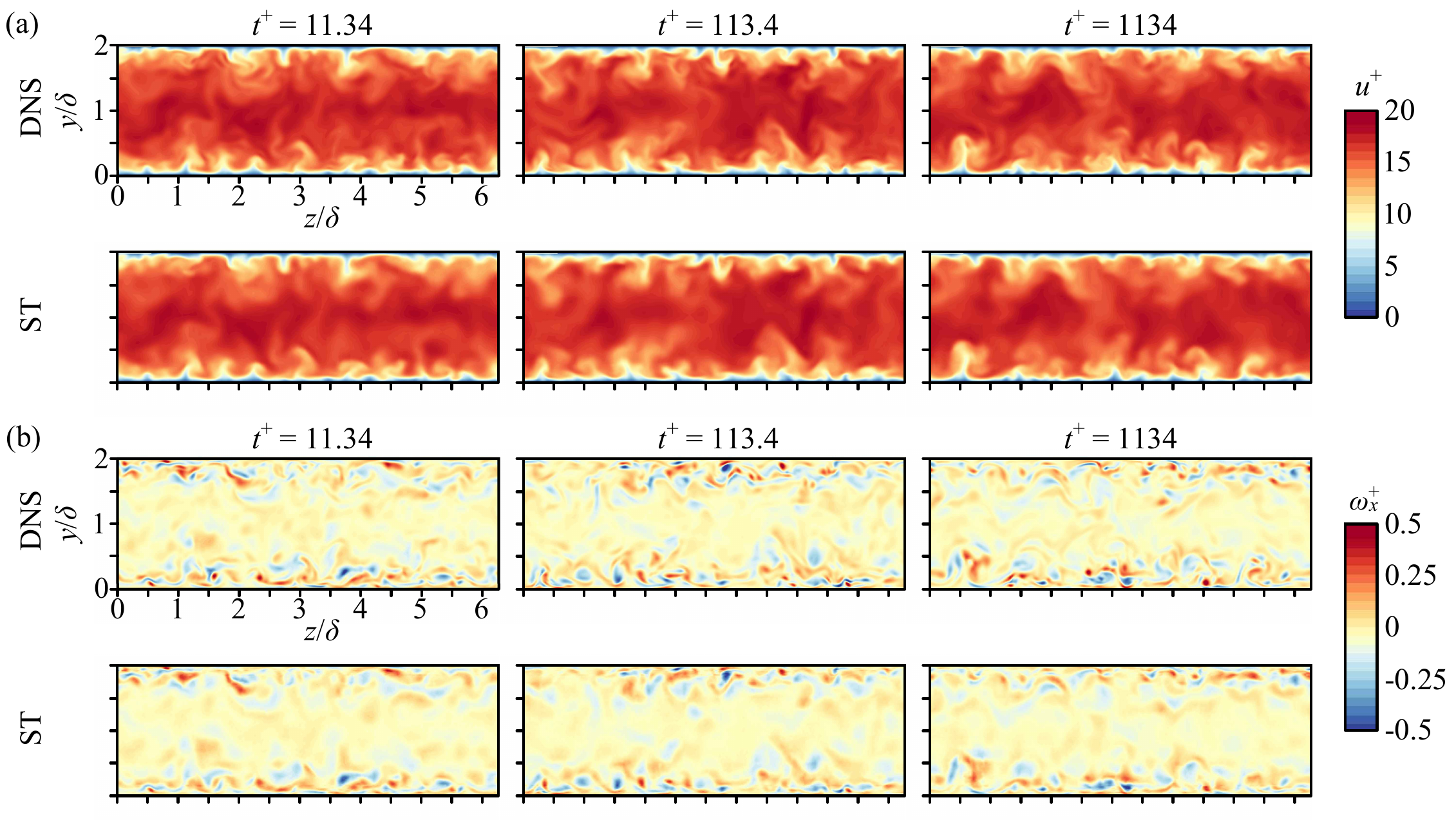}}
  \caption{Instantaneous streamwise (a) velocity field and (b) vorticity field for the case of turbulent channel flow at $Re_\tau$ = 180.}
\label{fig:F6}
\end{figure}

\begin{figure}[htbp]
  \centerline{\includegraphics[scale=0.4]{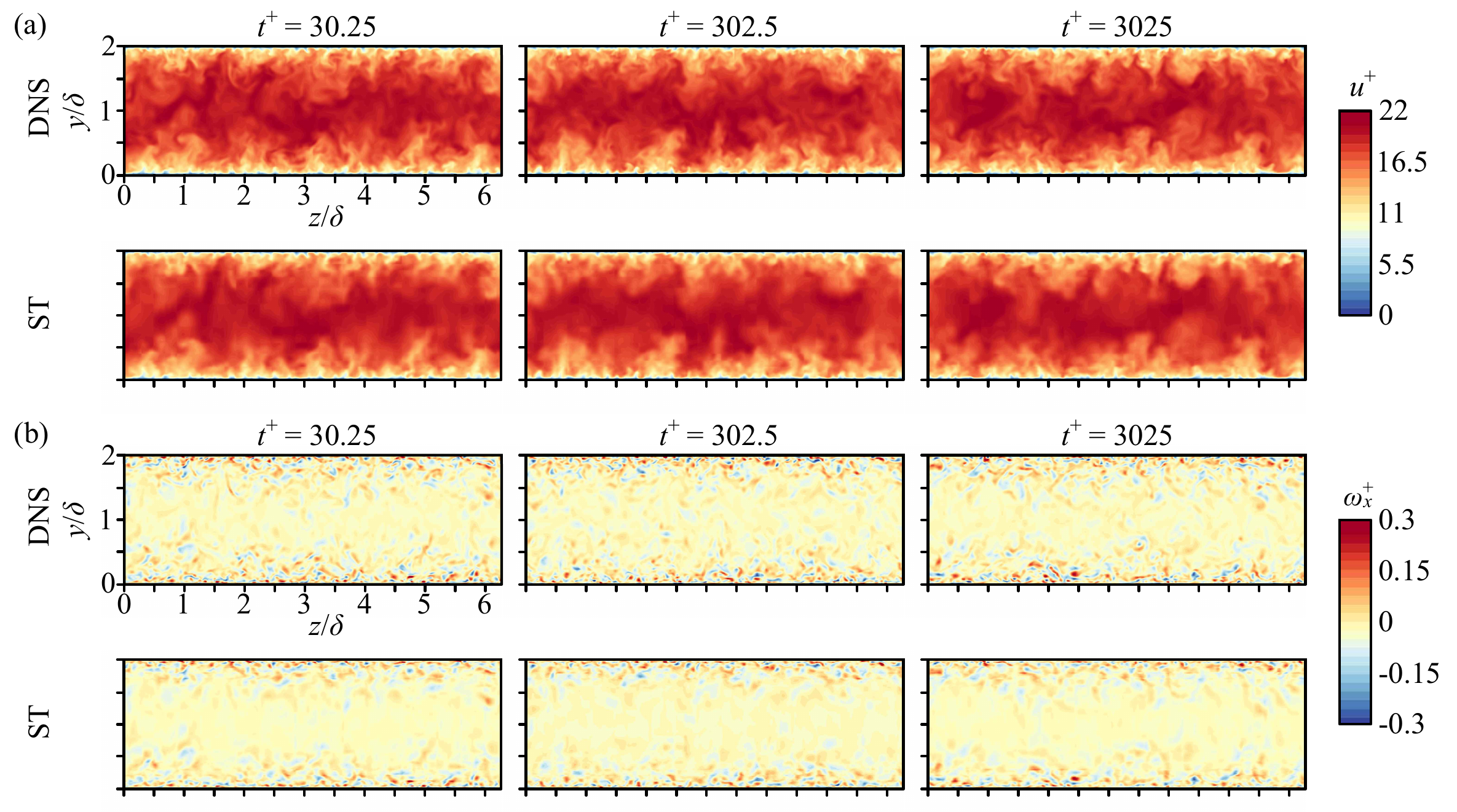}}
  \caption{Instantaneous streamwise (a) velocity field and (b) vorticity field for the case of turbulent channel flow at $Re_\tau$ = 550.}
\label{fig:F7}
\end{figure}

Figures~\ref{fig:F6} and \ref{fig:F7} display the instantaneous streamwise velocity field ($u^+$) and vorticity field ($\omega_x^+$) of the DNS and ST-decompressed results for three different time steps at $Re_\tau$ = 180 and $Re_\tau$ = 550, respectively. It can be observed that the ST model successfully compresses and decompresses the flow data at $Re_\tau$ = 180, yielding results that are consistent with the DNS data. Nonetheless, there are some visual disparities in the decompressed turbulent channel flow at $Re_\tau$ = 550, particularly in the representation of small-scale structures, while the dominant flow features and flow patterns have been well-preserved.

\begin{figure}[htbp]
  \centerline{\includegraphics[scale=0.3]{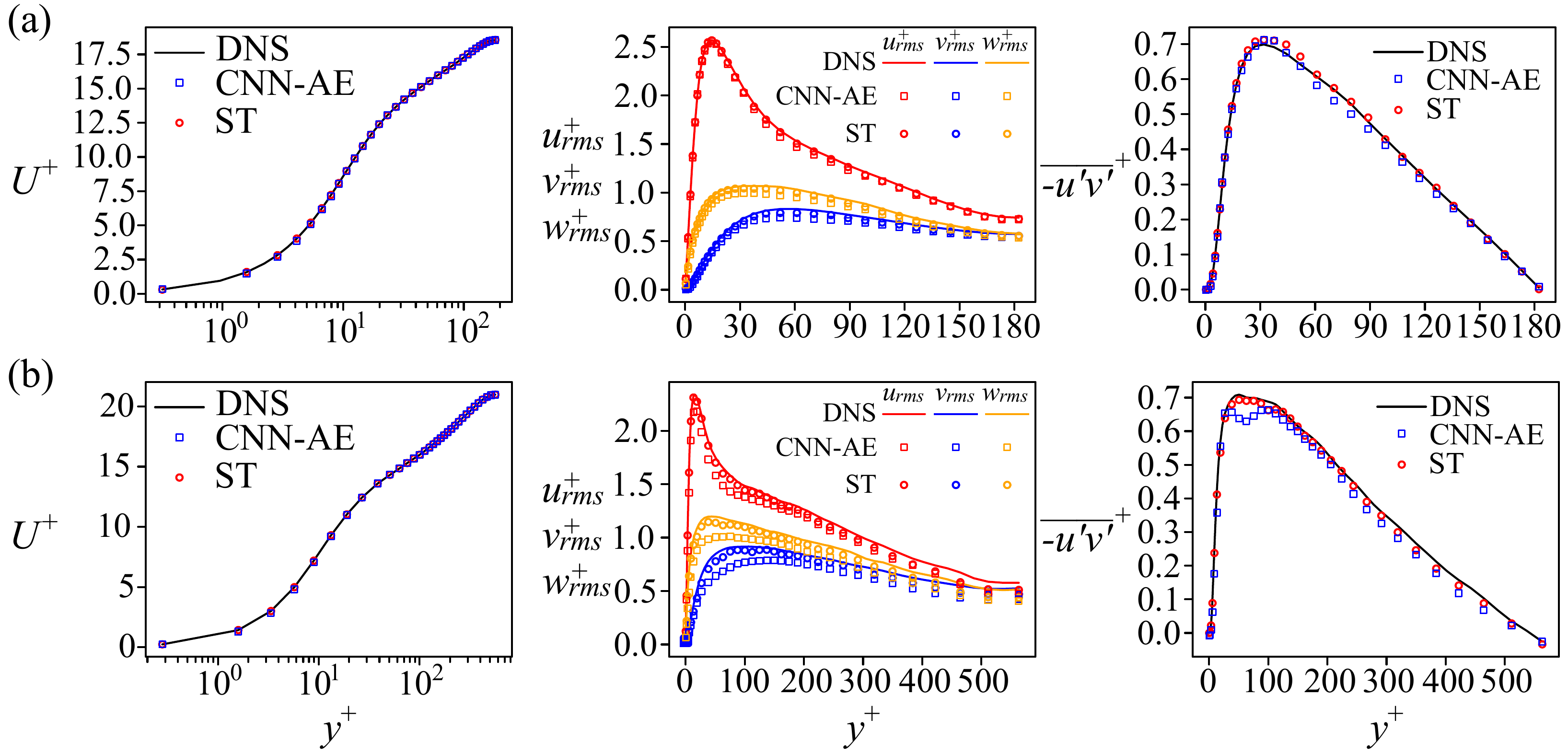}}
  \caption{Turbulent statistics for the turbulent channel flow at (a) $Re_\tau$ = 180 and (b) $Re_\tau$ = 550. Mean streamwise velocity profile (left), r.m.s. profiles for the three velocity components (middle), and Reynolds shear stress profile (right).}
\label{fig:F8}
\end{figure}

The turbulent statistics of the reconstructed velocity fields are compared with the turbulent statistics of the DNS turbulent channel flow at $Re_\tau$ = 180 and 550 in Figure~\ref{fig:F8} (a) and (b), respectively. The mean streamwise velocity ($U^+$) profiles of the decompressed flow using the ST model and the CNN-AE at $Re_\tau$ = 180 and 550 show accurate alignment with the profiles from the DNS data, covering the entire $y^+$ range. The comparison of the root-mean-square (r.m.s.) profiles of the velocity components ($u_{rms}^+$, $v_{rms}^+$ and $w_{rms}^+$) reveal a different observation. The r.m.s. profiles of the reconstructed flow obtained using the ST model fit well with the DNS data at both $Re_\tau$ = 180 and 550. In contrast, the CNN-AE produces relatively less accurate results, particularly for the flow at $Re_\tau$ = 550. Similarly, the Reynolds shear stress profile profiles have the same behavior as the r.m.s. profiles. This can be attributed to the fact that at higher $Re_\tau$, the flow becomes more complex and chaotic, making it more challenging for the CNN-AE to reconstruct the boundary region accurately.

\begin{figure}[htbp]
  \centerline{\includegraphics[scale=0.3]{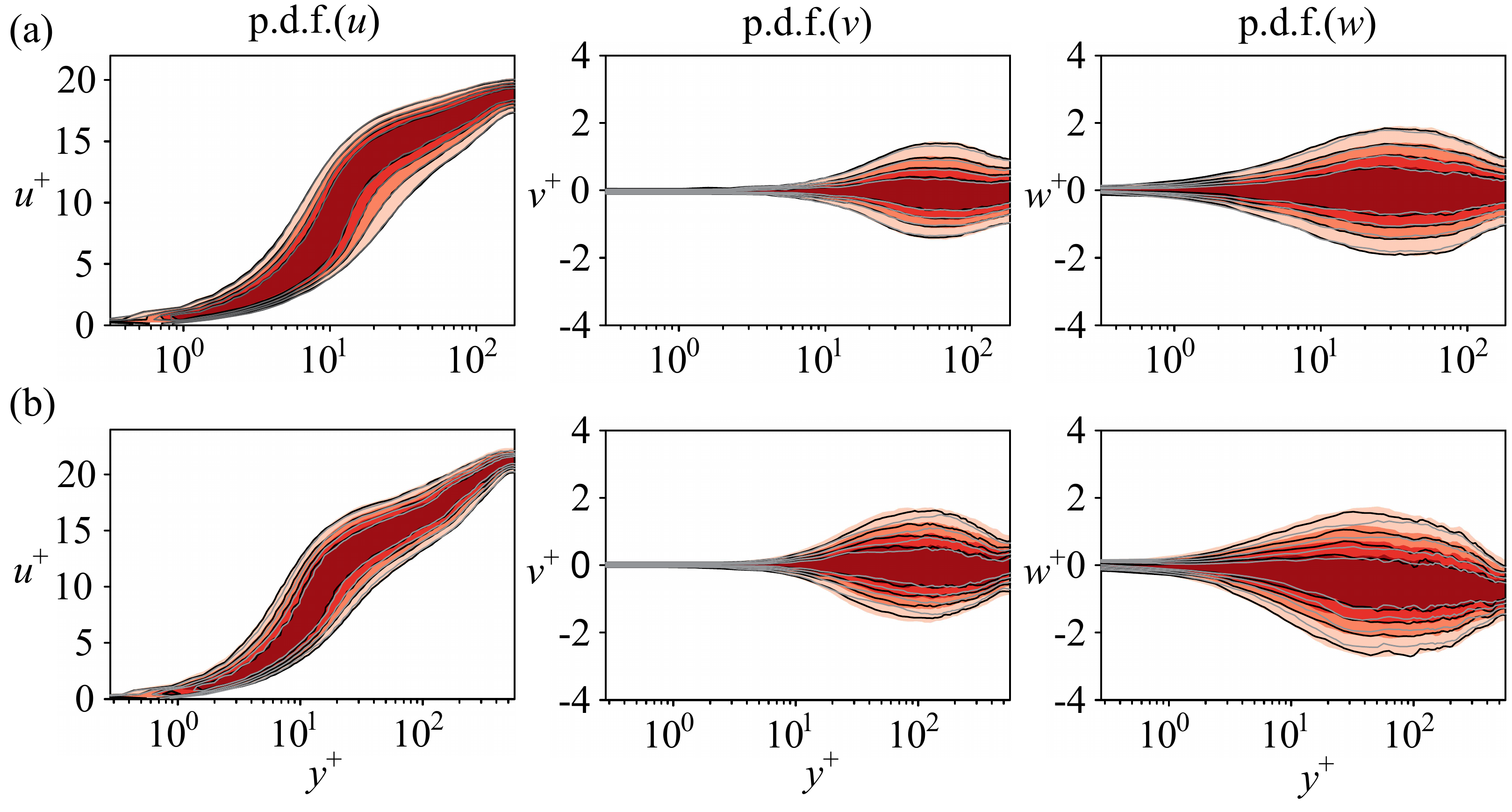}}
  \caption{Probability density function plots of the three velocity components (streamwise velocity on the left, wall-normal velocity in the middle, and spanwise velocity on the right) as a function of the wall-normal distance for the turbulent channel flow at (a) $Re_\tau$ = 180 and (b) $Re_\tau$ = 550. Shaded contours indicate the p.d.f. from the DNS data, while black contours and grey contours represent reconstruction results of the ST model and the CNN-AE, respectively. The contours levels are 20\%, 40\%, 60\% and 80\% of the maximum p.d.f.}
\label{fig:F9}
\end{figure}

The p.d.f. plots of the three velocity fields ($u^+$, $v^+$ and $w^+$) for $Re_\tau$ = 180 and 550 decompressed from the ST model and the CNN-AE are shown in Figure~\ref{fig:F9}. It can be observed that the p.d.f. of the reconstructed velocity components are consistent with the DNS results, while those from the CNN-AE exhibit a relatively high deviation, especially for the flow at $Re_\tau$ = 550. These results indicate that the ST model offers greater advantages in compressing and decompressing the flow data than the CNN-AE.

\begin{figure}[htbp]
  \centerline{\includegraphics[scale=0.3]{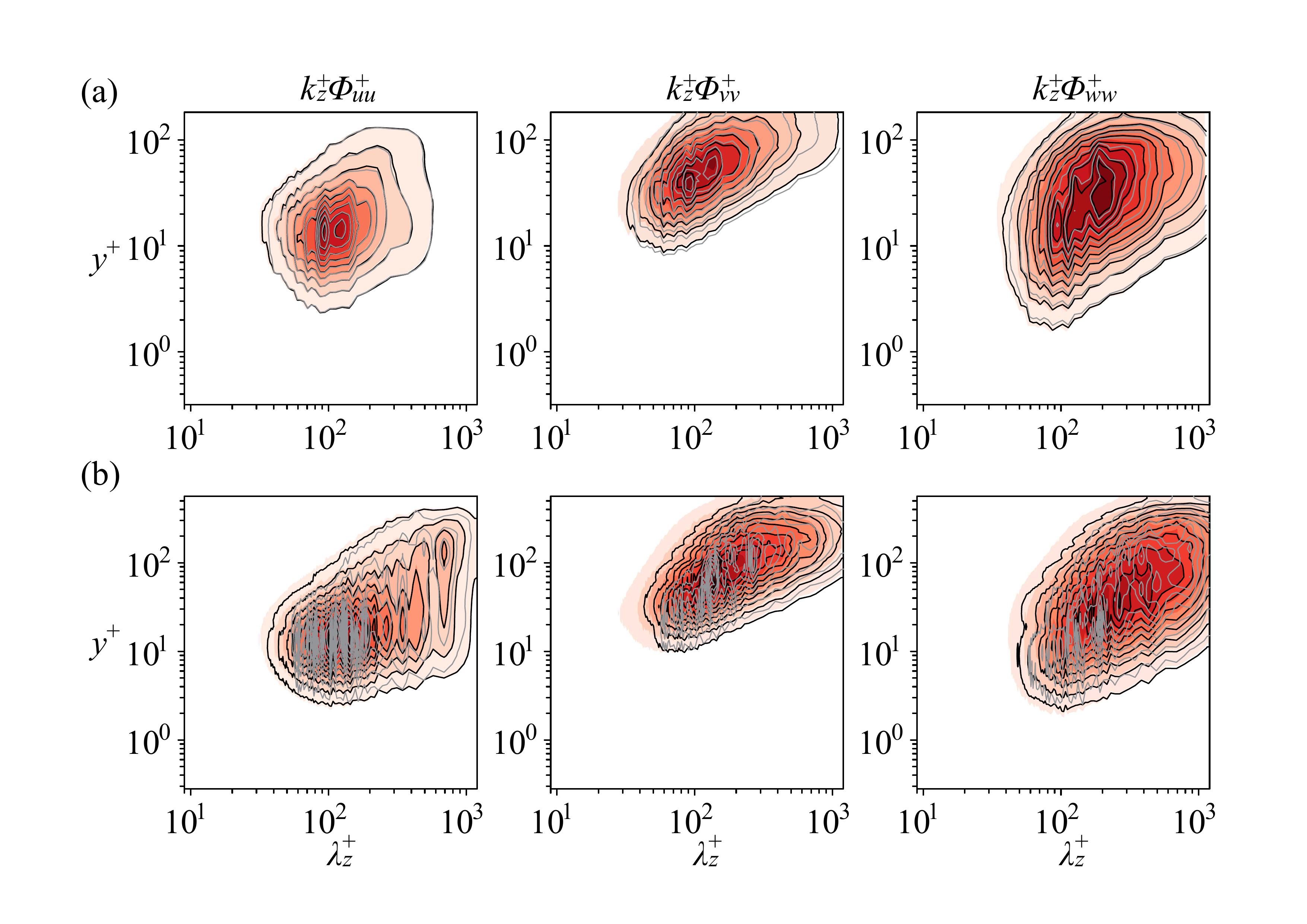}}
  \caption{Premultiplied spanwise wavenumber energy spectra of the three velocity components (streamwise velocity on the left, wall-normal velocity in the middle, and spanwise velocity on the right) as a function of the wall-normal distance and the spanwise wavelength for the turbulent channel flow at (a) $Re_\tau$ = 180 and (b) $Re_\tau$ = 550. Shaded contours show DNS data, while black and grey contours represent reconstruction results of the ST model and the CNN-AE, respectively. The contour levels are set at 10\% increments, ranging from 10\% to 90\% of the maximum premultiplied spanwise wavenumber energy spectra.}
\label{fig:F10}
\end{figure}

To further confirm the capability of the ST model in reconstructing genuine spatial spectra of the restored velocity fields, the premultiplied spanwise wavenumber energy spectra of the three velocity components denoted as $k_z \phi_{\xi \xi}$, are examined. Here, $\phi_{\xi \xi}$ denotes the spanwise wavenumber spectrum, $\xi$ means velocity component and $k_z$ is the spanwise wavenumber. Figure~\ref{fig:F10} shows the plots of the $k_z^+ \phi_{\xi \xi}^+$ as a function of the wall-normal distance $y^+$ and the spanwise wavelength $\lambda_z^+$. The spectra of the velocity components obtained from the ST model conform to the spectra from the DNS data with a small discrepancy observed at the high wavenumbers, while the $k_z^+ \phi_{\xi \xi}^+$ plots obtained from the CNN-AE are less accurate than those obtained from the ST model. These results further validate the ST model's outstanding ability to accurately capture the spatial distribution of the velocity fields.

\begin{figure}[htbp]
  \centerline{\includegraphics[scale=0.6]{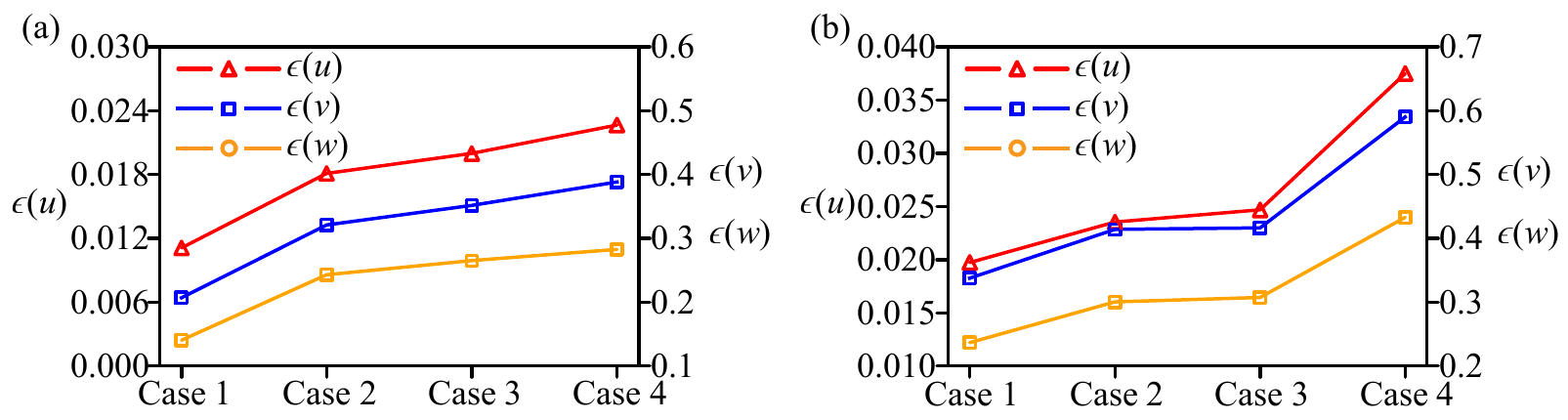}}
  \caption{Relative $L_2$-norm error of the decompressed velocity fields at (a) $Re_\tau$ = 180 and (b) $Re_\tau$ = 550. Cases 1, 2, and 3 correspond to the decompressed results from the ST with $CR$ = 16, 64, and 256, while Case 4 represents the decompressed results from the CNN-AE with $CR$ = 64.}
\label{fig:F11}
\end{figure}

The compression and decompression accuracy of the ST model and the CNN-AE at $Re_\tau$ = 180 and 550 are investigated by using the $L_2$-norm relative error of the velocity fields:
\begin{equation} \label{eqn:eq14}
\epsilon(\xi) = \frac{1}{I} \sum_{i=1}^{I} \frac{\lVert \hat{\xi}_{i} - \xi_{i} \lVert_2}{\lVert \xi_{i} \lVert_2}.
\end{equation}
\noindent where $\hat{\xi}_{i}$ and $\xi_i$ denote the decompressed velocity fields by each model and the DNS data, respectively. $I$ represents the total number of test snapshots, which is set to 1,000. Figure~\ref{fig:F11} presents the $L_2$-norm relative error for the reconstructed flow at (a) $Re_\tau$ = 180 and (b) $Re_\tau$ = 550. As shown, the ST model achieves lower errors than the CNN-AE for the two Reynolds numbers with the same $CR$, indicating the superior performance of the ST model. These results further confirm that the ST model outperforms the CNN-AE. This can be attributed to the ability of the ST model to capture long-distance spatial correlation, making it more suitable for non-uniformly distributed data. These results give confidence that the ST model can be applied to complex geometric flow data such as pipe flow by adjusting the window segmentation strategy and masking mechanism, while for the CNN-AE, the use of the padding operation can result in significant errors at the boundaries.

In addition to the $CR$ = 64 mentioned earlier in this section, here, two more $CR$ values are added to validate the ability of the ST model. The errors of the three velocity components increase relatively as the $CR$ increases, which aligns with the trade-off between $CR$ and reconstruction quality. However, even with $CR$ = 256, the error of the ST model is still smaller than the result of the CNN-AE with $CR$ = 64, which indicates that the proposed ST model is robust for different compression ratios. Moreover, the decompression flow exhibits larger errors at the high $Re_\tau$, which is attributable to the increased turbulence and complexity of the flow field at higher Reynolds numbers. Nevertheless, Figure~\ref{fig:F11} (b) shows that the error of the proposed model does not increase significantly with increasing $CR$, demonstrating that our model can still achieve high accuracy even for challenging recovery cases.

Notably, the transfer learning (TL) technique \cite{Yousifetal2021, Yousifetal2022b} is employed in this study to decrease the training time by leveraging the weights of a trained model to initialize another model. The ST model is first trained on the turbulent channel flow at $Re_\tau$ = 180. Subsequently, the weights of the trained ST model are transferred to initialize the model for the turbulent channel flow at $Re_\tau$ = 550, thus enabling faster convergence. The results indicate that TL can reduce the training time by 64\% without compromising the accuracy, and the reduced training time is relatively greater than that reported in Yousif {\it et al.} \cite{Yousifetal2023b} since this study did not reduce the amount of training data.  

Finally, it is important to consider the computational cost of the ST model. When $CR$ = 64, the ST model has a total of approximately $6.60\times 10^6$ trainable parameters ($3.30\times 10^6$ for the encoder part and $3.30\times 10^6$ for the decoder part). When training the ST model for turbulent channel flow at $Re_\tau$ = 180 and 550, it takes around 40 and 14 hours, respectively, using a single NVIDIA TITAN RTX GPU with the aid of TL. Despite the relatively long training time, the computational cost is a one-time expense. After the model training is completed, the computational cost of compressing and decompressing flow data is negligible, which meets the requirements for fast and efficient data processing.

\section{Conclusions}\label{sec:Conclusions}
This study proposed an efficient compression deep-learning method for turbulent data storage and transmission using a Swin-Transformer-based model, called the ST model. A physical constraints-based loss function was made of the velocity gradient error, Reynolds stress error, energy spectrum error, and velocity error, which guides the model's learning process to capture the underlying physical behavior of the turbulent flow. 

First, the forced isotropic turbulent flow at $Re_\lambda$ = 418 obtained from the JHTDB was considered as a demonstration case of the ST model's ability to compress and decompress the turbulent data. The instantaneous and statistical results of the isotropic flow exhibit the outstanding capability of the ST model to compress large data for storage and transmission and restore it with factual information. Furthermore, the ability of the ST model was tested and validated by the turbulent channel flow at $Re_\tau$ = 180 and 550 generated by DNS. The restored instantaneous velocity fields showed excellent results that matched well with the DNS data. In addition to visual analysis, the statistical analysis of the velocity fields also yielded accurate results, with the exception of a minor deviation in the flow at $Re_\tau$ = 550, which can be attributed to the increased chaotic nature of the turbulence with the increasing of Reynolds number. The probability density function and the premultiplied spanwise wavenumber energy spectra agreed with the ground truth data, indicating the accurate spatial distribution of the reconstructed velocity fields. While the above results were obtained using $CR$ = 64, a higher $CR$ = 256 was used to prove the robustness of the ST model to the change in the $CR$. The relative error plots denoted that the errors remained significantly low even under the high compression ratio, confirming the reliable compression power of the model.

In addition, the proposed ST model was compared in terms of performance with a CNN-AE. The statistical profiles of the turbulent channel flow revealed that the results from the ST model were significantly more consistent with the DNS data than those obtained by the CNN-AE, indicating the superior ability of the ST model to compress and decompress the turbulent flow. The comparisons of p.d.f. and the energy spectra further supported the ST model's superior ability, especially for the turbulent channel flow at $Re_\tau$ = 550. Moreover, the relative error of the CNN-AE was much higher than the ST model under the same $CR$. All the compared results suggested that the ST model can achieve better restoration than the CNN-AE for non-uniform flow data. Finally, the effect of transfer learning that leverages the weights of a trained model to initialize another model was checked by transferring the weights of the trained ST model for the flow at $Re_\tau$ = 180 to initialize the model for the flow at $Re_\tau$ = 550. The results showed that TL reduced the training time by 64\% without diminishing the correctness.

In this study, the ST model combined with a physical constraints-based loss function provides a powerful data compression and decompression solution in fluid mechanics, which can provide high compression ratios and accurate results. This can result in reduced data storage and transmission requirements and consequently can increase the efficiency of data-driven turbulence research.

\begin{acknowledgments}
This work was supported by 'Human Resources Program in Energy Technology' of the Korea Institute of Energy Technology Evaluation and Planning (KETEP), granted financial resource from the Ministry of Trade, Industry \& Energy, Republic of Korea (no. 20214000000140). In addition, this work was supported by the National Research Foundation of Korea (NRF) grant funded by the Korea government (MSIP) (no. 2019R1I1A3A01058576). 
\end{acknowledgments}

\section*{Data Availability}
The data that supports the findings of this study are available within this article.

\bibliography{my-bib}

\end{document}